\documentclass{article}

\usepackage{graphicx}
\usepackage{natbib}
\usepackage{amssymb}

\usepackage{my_definitions}

\title
{Are Lagrangian stochastic models at odds with statistical theories of
relative dispersion?}

\author{Alberto Maurizi\\
ISAC-CNR, via Gobetti 101, 40129 Bologna, Italy
}

\begin{document}

\maketitle

\begin{abstract}
In an article on statistical modelling of turbulent relative dispersion,
\citet[p. 402]{franzese_etal-jfm-2007} commented on Lagrangian stochastic
models and reported some concern about the consistency between statistical
and stochastic modelling of turbulent dispersion.
In this short article, comparison of the two approaches is performed. As far
as the dependence of models from turbulence constants is concerned, the two
theoretical approaches are found to be in perfect agreement eliminating every
possible concern.
\end{abstract}

\section{Introduction}
In an article on statistical theory of relative dispersion,
\citet{franzese_etal-jfm-2007} (hereinafter FC) found that within their
approximations, the Richardson constant $C_r$ is expressed by
\begin{equation}
C_r=6\alpha C_0\,,
\label{eq:g-FC}
\end{equation}
where $C_0$ is the, supposedly universal, Kolmogorov constant of the
second-order Lagrangian structure function and $\alpha$ is expressed by
\begin{equation}
\alpha=\frac{C_L}{2}\left[\left(1+\frac{4}{3C_L}\right)^{1/2}-1\right]^3\,,
\label{eq:alpha}
\end{equation}
where $C_L$ is a measure of the ratio between ``a length scale of the energy
containing eddies'' (FC) and $\sigma_u T_L$, $\sigma_u$ being the r.m.s of
the turbulent velocity and $T_L$ the Lagrangian integral time scale. $C_L$ is
then determined by some closure assumption and turns out to be 8/3. Using
arguments based on their definition of scales, FC conclude that:
\begin{quote}
``The anomalous inverse relation between $C_0$ and $C_r$ observed in
stochastic Lagrangian models [\ldots] arises from the violation of (4.11)
[their numbering\footnote{Actually, an equation similar to \eqref{eq:beta} in
the present paper.}]. Increasing $C_0$ with a fixed $C_k$ determines a
spurious increase in $C_\sigma$, namely, the proportion between Eulerian and
Lagrangian scales is altered, with an overestimated value of
$\mathcal{L}_E$\footnote{The same as $\lambda_E$ in this context.}. In such
conditions, the particles separate at a slower rate because the fraction of
energy used for the separation process is underestimated.''
\end{quote}
It is not clear whether the FC claim is that Lagrangian stochastic
models (LSM) are incorrect while statistical theories, by contrast, display
the ``correct'' features. Nevertheless, a clarification on the connection
between these two approaches can be worthwhile in the light of the doubt
generated by the lack of clarity of the FC article on this particular aspect.
In fact the comparison of Lagrangian statistical and stochastic models was not
the main purpose of FC. Thus, in the present article the opportunity is
taken to clarify the problem in the framework defined by
\citet{maurizi_etal-pre-2004} (hereinafter MPT).

\section{A reminder and a critical analysis of MPT results}
It is useful here to recall the basics of the MPT work. The idea was to
study some general properties of the WM models for relative turbulent
dispersion. The frame is that of \citet{kolmogorov-1941} theory (hereinafter
K41) and the approach is the Well Mixed (WM) condition
\citep{thomson-jfm-1987,thomson-jfm-1990}. The consideration on which the
MPT analysis is based, is the fact that Eulerian and Lagrangian properties
co-exist in the description of relative dispersion and that their typical
scales (Eulerian length and Lagrangian time scales) can be used to highlight model
properties.

From K41 it is known that in homogeneous isotropic turbulence, the
second-order longitudinal Eulerian structure function $S^{(2)}_E$ in the
inertial sub-range is:
\begin{equation}
S^{(2)}_E
\equiv \mean{ \{ [\mathbf{u}(\mathbf{x}+\Delta\mathbf{r})-\mathbf{u}(\mathbf{x}) ] \cdot
\Delta\mathbf{r}(\Delta r)^{-1} \}^2}
=C_k (\varepsilon \Delta r)^{2/3}
\label{eq:S2E}
\end{equation}
for $\eta \ll \Delta r\equiv||\Delta \mathbf{r}||\ll L_E$, where $varepsilon$
is the turbulent kinetic energy dissipation rate, $\eta$ is the
Kolmogorov microscale and $L_E$ is the Eulerian integral length scale. Using the
relationship between second-order structure function and correlation
coefficient $S^{(2)}(\Delta)=2\sigma_u^2[1-R(\Delta)]$ (where $\Delta$ is
either $\Delta r$ or $\Delta t$ for Eulerian and Lagrangian formulation,
respectively) it turns out that in the inertial subrange
\begin{equation}
R_E(\Delta r)
=
1-\frac{S^{(2)}_E}{2\sigma_u^2}
=
1-\frac{C_k(\varepsilon \Delta r)^{2/3}}{2\sigma_u^2}\,.
\label{eq:RE}
\end{equation}
\Eqref{eq:RE} can be used as a definition for a length scale
\begin{equation}
\lambda_E
=\left(\frac{2}{C_k}\right)^{3/2}\frac{\sigma_u^3}{\varepsilon}\,.
\label{eq:lambda}
\end{equation}

It is straightforward to follow the same procedure in the Lagrangian frame. In
fact, on dimensional grounds it is known that \citep{monin_yaglom-1975} at the
leading order in $\Delta t$,
\begin{equation}
S^{(2)}_{L_{i}}
\equiv \mean{[v_i(t+\Delta t)-v_i(t)]^2}
=C_0 (\varepsilon \Delta t)
\label{eq:S2L}
\end{equation}
in which $v_i(t)=u_i(X(t),t)$ is the Lagrangian velocity, \ie the Eulerian
velocity at the particle position $X(t)$. \Eqref{eq:S2L} is valid for
$\tau_\eta \ll \Delta t\ll T_L$, $\tau_\eta$ being the Kolmogorov time scale.
This gives for the Lagrangian autocorrelation function:
\begin{equation}
R_{L_{i}}(\Delta t)
= 1-\frac{C_0(\varepsilon \Delta t)}{2\sigma_u^2}
\label{eq:RL}
\end{equation}
that can be used as a definition for a time scale
\begin{equation}
\tau_L=\frac{2\sigma_u^2}{C_0\varepsilon}\,.
\label{eq:tau}
\end{equation}
\Eqref{eq:tau} represents the Lagrangian counterpart of \Eqref{eq:lambda} and
corresponds to the known relationship given by \citet{tennekes-1982}.

It can be observed here that the above definitions link the scales to their
corresponding constants: $C_k$ and $C_0$ for the Eulerian length and
Lagrangian time scales, respectively.
It is worth pointing out that the presence of redundant scales (length, time,
velocity) is not surprising considering that they are in fact the scales of
the independent ingredients of a LSM: the Lagrangian structure function enters
for compatibility with small scale behaviour \citep{thomson-jfm-1987}; the
two-point Eulerian structure function is imposed by the WM condition
\citep{thomson-jfm-1990} through the Eulerian probability density function
(pdf) of the flow velocity, and the kinetic energy is not directly connected to
$\lambda_E\tau_L^{-1}$ and therefore it is another (independent) parameter of
the Eulerian pdf. This redundancy can be regarded as the manifestation of the
competing role of Eulerian and Lagrangian scales in relative dispersion.
``Real'' turbulence does not display any variability of the constants because
Eulerian and Lagrangian properties are both uniquely determined by dynamical
equations.
Note also that LSM theory is valid for infinite Reynolds number Re and
therefore no variations of constants can be attributed to variations in Re.
However, varying constants is possible in models where Eulerian and Lagrangian
properties are imposed as ``phenomenological'' model constraints.

The above defined scales can be used to render non-dimensional the
Fokker-Planck equation for the probability density function of the process
$p=p(\mathbf{u},\mathbf{x};t)$, where
$\mathbf{u}\equiv(\mathbf{u}^{(1)},\mathbf{u}^{(2)})$ and
$\mathbf{x}\equiv(\mathbf{x}^{(1)},\mathbf{x}^{(2)})$ (with superscript
referring to particle 1 and 2):
\begin{equation}
\frac{\partial}{\partial t} p + \beta \frac{\partial}{\partial x_i} (u_i p) +
\frac{\partial}{\partial u_i} (a_i p) = \frac{\partial^2}{\partial
{u_i}\partial u_i} p
\label{eq:FPE}
\end{equation}
where all the quantities involved are non-dimensional: $t\rightarrow\tau_L t$,
$x_i\rightarrow\lambda_E x_i$, $u_i\rightarrow\sigma_u u_i$,
$p\rightarrow\sigma_u^{-3}p$ and
$a_i\rightarrow\sigma_u\tau_L^{-1}a_i$.
With these scalings, the constant
$\beta$ is the sole remaining parameter of \Eqref{eq:FPE} and is
expressed by
\begin{equation}
\beta=\frac{\sigma_u\tau_L}{\lambda_E}\equiv\left(\frac{C_k^3}{2C_0^2}\right)^{1/2}
\label{eq:beta}
\end{equation}
which is a non-dimensional combination of the above-defined scales and can be
recognised to be a possible definition for the the quantity commonly known as
Lagrangian-to-Eulerian scale ratio.
It can be observed that the alternative choice
$a_i\rightarrow\sigma_u^2\lambda_E^{-1}a_i$ for the drift term scaling is
still possible but, while changing the form of \Eqref{eq:FPE}, would not
affect its dependence on $\beta$ as the unique parameter.

The results of MPT are worth a comment. The arguments used are, in general,
not sufficient to state that ``any'' solution of \Eqref{eq:FPE} depends solely
on $\beta$ because, in fact, the drift term $a_i$ results from the application
of the WM condition:
\begin{equation}
a_i=\frac{C_0\varepsilon}{2}\frac{\partial}{\partial u_i} \log P_E +
\frac{\Phi_i}{P_E} \,,
\end{equation}
where
\begin{equation}
\frac{\partial}{\partial u_i}\Phi_i=\frac{\partial}{\partial
t}P_E+u_i\frac{\partial}{\partial x_i} P_E \,,
\end{equation}
and in general can depend also on other parameters via $P_E$ and/or via the
assumptions made to remove the indeterminacy intrinsic to the WM condition.

Nevertheless, it can be demonstrated that for Gaussian $P_E$
\citep{thomson-jfm-1990,borgas_etal-jfm-1994}, the non-dimensional $a_i$
actually depends solely on $\beta$ confirming the MPT statement for this class
of models.  In fact, considering the general class of of Gaussian models
presented by \citet{borgas_etal-jfm-1994}, the (dimensional) drift term always
has a form of the type $a_i=\sigma_u\tau_L^{-1} A_i + \sigma_u^2\lambda_E^{-1}
B_i$ from which $\sigma_u^{-1}\tau_L a_i = A_i + \beta B_i$ with $A_i$ and
$B_i$ non-dimensional and independent of $C_k$ and $C_0$.

Full 3-dimensional (3D) solutions are presently beyond reach. However,
quasi--one-dimensional (Q1D) approach \citep{kurbanmuradov-mcma-1997} makes
it possible to estimate to what extent the MPT conclusions are valid also
for departures of $P_E$ from Gaussianity.
In \citet{kurbanmuradov-mcma-1997} a systematic study of the behaviour of the
Q1D model in response to variation of non-Gaussian properties was carried out.
The result was that the differences observed as a result of variations of
$C_0$ (\ie $\beta$) are much larger than those that result from departure from
Gaussianity. In addition, \citet{kurbanmuradov-mcma-1997} also noticed that
non-Gaussian and Gaussian Q1D models behave qualitatively the same.

It can be inferred that, at least in the Q1D frame, $\beta$ is still the
driving parameter of the non-dimensional Fokker-Plank solutions, with
non-Gaussianity playing a minor role. In other words, the drift term $a$
can be expressed as $f(\beta,G)=f_0(\beta) + O(G)$ with $G$ being the parameter
driving the non-Gaussianity of $P_E$.

Another property that is expected to play a role so as to introduce a further
parameter, is rotation \citep{sawford-blm-1999} which is related to the
non-uniqueness problem. However, investigating also on the consequences of
this aspect is beyond the scope of the present work.

\section{Connection between Lagrangian Statistical and Stochastic approaches}
In terms of the T90 theory, the validity of MPT results, although
rigorously true only for Gaussian (but still approximately true for
non-Gaussian $P_E$), means that once the non-uniqueness problem is resolved
(by selecting one of the infinitely many solutions of the WM conditioned
problem) results of \Eqref{eq:FPE} depend solely on $\beta$.

Considering the Richardson law
\begin{equation}
\mean{\Delta X_i^2}=C_r \varepsilon t^3
\label{eq:R26}
\end{equation}
reducing it to non-dimensional form, taking into account that $\Delta X_i$ is
a Lagrangian quantity, \ie $\mean{\Delta
X_i^2}\rightarrow\sigma_u\tau_L^{-1}\mean{\Delta X_i^2}$
\citep{maurizi_etal-blm-2006}\footnote{It is straightforward to show that,
even using the scaling $\mean{\Delta
X_i^2}\rightarrow\sigma_u^2\lambda_E^{-1}\mean{\Delta X_i^2}$ does not affect
the present arguments but only the form of \Eqref{eq:g-MPT}.}, it turns out
that
\begin{equation}
\mean{\Delta X_i^2}=2C_r^* t^3
\label{eq:R26-non-dim}
\end{equation}
where all the variables are non-dimensional and $C_r^*=C_r C_0^{-1}$ is the
normalised Richardson coefficient.

\begin{figure}
\includegraphics{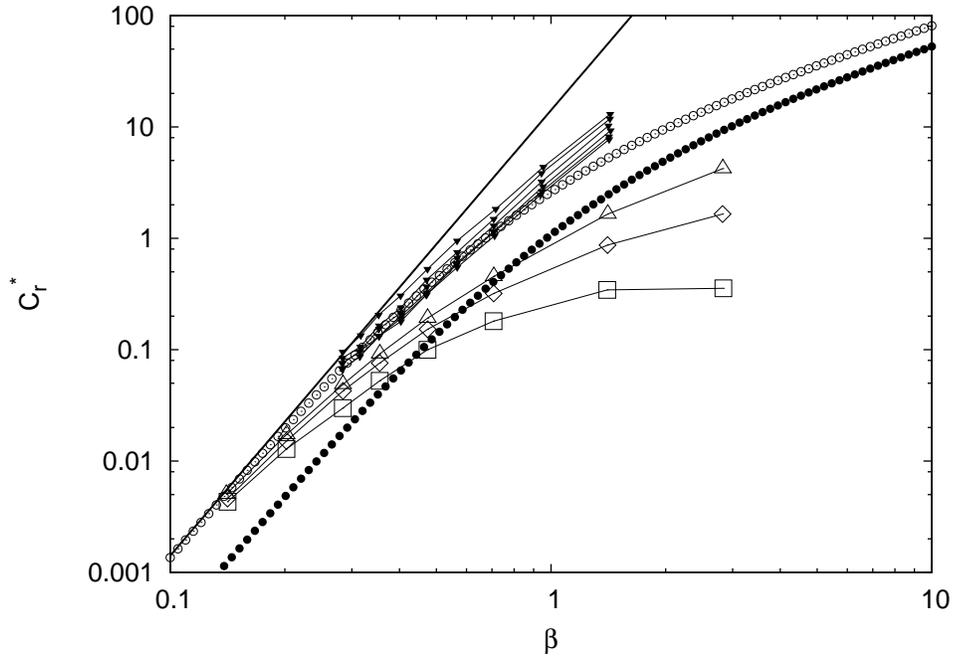}
\caption{Normalised Richardson coefficient $C_r^*$ as a function of the
Lagrangian-to-Eulerian scale ratio $\beta$.
Symbols are as follows:
open square, \citet[model 4.2a]{borgas_etal-jfm-1994};
open diamonds, \citet[model 7.6 with $\varphi=-0.4$]{borgas_etal-jfm-1994};
open triangle, \citet[model 4.3]{borgas_etal-jfm-1994};
small reverse full triangle, \citet{kurbanmuradov-mcma-1997} for different
departures from Gaussianity.
Continuous line is the diffusion limit \citep{borgas_etal-jfm-1994}.
Line dotted with full circles is \Eqref{eq:g-FC} (FC) with $C_L=0.53\beta^{-2}$
and line dotted with open circles is the same equation with
$C_L=0.25\beta^{-2}$, the value for consistency with BS94 as
$\beta\rightarrow0$.
}
\label{fig:g-star}
\end{figure}

This result was used to analyse and to arrange systematically data from
literature. \Figref{fig:g-star} reports results from different LSM both
Gaussian \citep{borgas_etal-jfm-1994} and non-Gaussian
\citep{kurbanmuradov-mcma-1997}. The regularity of the behaviour of $C_r^*$
with varying $\beta$ is striking. All the models are in agreement with the
diffusion limit \citep{borgas_etal-jfm-1994} and their rate of growth with
$\beta$ is also monotonic. Moreover, as anticipated above,
departures from Gaussianity do not modify the general picture given by
MPT.

A direct consequence of \Eqref{eq:R26-non-dim} is that if $\mean{\Delta
X_i^2}$ is solution of a given WM model it will depend on $\beta$ only so as
$C_r^*$, and consequently
\begin{equation}
C_r=F(\beta)C_0
\label{eq:g-MPT}
\end{equation}
where the functional form of $F(\beta)=0.5\mean{\Delta x^2}t^{-3}$ depends
only on the assumption made to close the non-uniqueness problem.
It is clear now that \Eqref{eq:g-FC} and \Eqref{eq:g-MPT} are equivalent when
viewed in the frame of the scaling described so far proving the qualitative
consistency of the Lagrangian statistical and stochastic (WM) approaches.

While it is clear that there is a dependence of the WM solution on $\beta$, it
is not evident if FC.
In fact, finding a dependence of $\alpha$ on $\beta$ in the FC theory
development is not straightforward because
a closure assumption is used
before its explicit definition so that the dependence of
$\alpha$ on $\beta$ is hidden.

In their Section 5, FC state that
\begin{equation}
T_y/\tau_L=(3\alpha/4)^{-1/3}
\label{eq:Ty}
\end{equation}
where $T_y$ is defined as the time at which the cloud of marked particles
reaches the dimension at which it starts to behave diffusively.
However, \Eqref{eq:Ty} is a consequence of the closure assumption made in
their subsequent Section 7. Thus before introducing the assumption that then
leads to $C_L=8/3$,
the actual expression of $C_L$ reads:
\begin{equation}
C_L=\frac{4}{3}\left[\left(\frac{\tau_\mathrm{L}}{T_y}+1\right)^2-1\right]^{-1}
\label{eq:C_L}
\end{equation}
and consequently, from \Eqref{eq:alpha}, the following expression holds:
\begin{equation}
\alpha=\frac{2}{3}\frac{(\tau_L/T_y)^2}{(\tau_L/T_y)+2}\,.
\label{eq:alpha-revisited}
\end{equation}

It can be argued that $T_y$ must be a function of the
Lagrangian-to-Eulerian scale ratio $\beta$.
Evidence for this dependence comes from asymptotic behaviour in the ideal
limiting cases: in the limit of infinite spatial correlation
($\beta\rightarrow 0$), particles do not separate so that
$T_y\rightarrow\infty$, while for vanishing spatial correlation
($\beta\rightarrow\infty$) the two-particles are independent since the
beginning so that $T_y\rightarrow \tau_L$.

Although the above arguments clearly indicate that $T_y/\tau_L$ must be a
function of $\beta$, it is impossible to proceed in this direction without further
assumptions.
However, any arbitrary assumption can be avoided noting that, being defined as the ratio
between a measure ``of a length scale of the energy containing eddies'' and
$\sigma^2\tau_L^{-1}$, $C_L$ turns out to be proportional to $\beta^{-2}$.
This consideration forces to recognise that the quantity $6C_\sigma$ in FC is
inessential and can be set to unit. In fact it is the ratio between two
quantities both proportional to $\mean{u^2}^3\varepsilon^{-2}$.

Substituting the relationship between $C_L$ and $\beta$ in
\Eqref{eq:alpha}, gives
\begin{equation}
\alpha=\frac{\gamma}{2\beta^2}\left[\left(1+\frac{4\beta^2}{3\gamma}\right)^{1/2}-1\right]^3
\label{eq:my-alpha2}
\end{equation}
with $\gamma$ to be determined.
\Eqref{eq:my-alpha2} can be used to exploit the exact dependence of
\Eqref{eq:g-FC} from $\beta$. Using the FC values: $\alpha=(18\sqrt{6}-44)/6$
and $\beta=0.44$, it turns out that $\gamma\simeq0.53$. The curve representing
FC model in terms of $g^*$ as a function of $\beta$ is reported in
\Figref{fig:g-star} for comparison with LSM results.

It can be observed that \Eqref{eq:my-alpha2} for $\beta\rightarrow0$, shows
the same power law dependence ($\beta^4$) as the BS94 limit for Gaussian LSM,
while having a very different coefficient as can be appreciated in
\Figref{fig:g-star}.
In fact, using the BS94 limit as constraint for \Eqref{eq:my-alpha2}
gives $\gamma\simeq0.25$ which, in terms of the FC closure, means that instead
of $T_y\simeq2.22\tau_L$\footnote{This is the value to be used in the exact
Ornstein-Uhlenbeck solution to obtain the FC value $C_L=8/3$ that, in turn, is
obtained by using $T_y=2$ with an approximate solution.} one should use
$T_y\simeq1.38\tau_L$. With this ``BS94 compliant'' closure, for $\beta=0.44$,
it results that $g^*\simeq0.31$ which is more than three times larger that the
value given by FC $g^*=0.09$. This highlight a strong sensitivity of FC results
to the value selected for the closure. The resulting curve is reported in
\Figref{fig:g-star}.

\section{Conclusions}
The analysis performed shown that the proportionality between $C_r$ and
$C_0$ is common to both Lagrangian statistical and stochastic derived models once,
using inertial sub-range scaling, the Lagrangian-to-Eulerian scale ratio
$\beta$ is recognised as driving parameter, and then kept constant. There is no
intrinsic violation of this scale ratio in LSM in that both $C_K$ and $C_0$
can be varied independently.

It was also shown that the connections between the two approaches is even more
intimate in that also the results of FC model formally depend on $\beta$.
In addition, FC model was shown to depend strongly on the value adopted for
the transition time from ballistic to diffusive regime which is a
rather poorly definable quantity.

In view of the results presented here, the consistency between Lagrangian
theories is not surprising at all because the ingredients used for both
approaches are the same and both rely on K41. Moreover, being the
results of FC derived from a pure statistical theory \textit{\`a la}
\citet{batchelor-1952}, its success and the consistency with
the LSM, enhances the validity of the latter rather than invalidates it.

\section*{acknowledgements}
The author would like to thank Gianni Pagnini, Brian Sawford and Francesco
Tampieri for useful discussions and comments and for the revision of the
manuscript.

\bibliographystyle{personal}
\bibliography{export}

\begin{thebibliography}{12}
\expandafter\ifx\csname natexlab\endcsname\relax\def\natexlab#1{#1}\fi

\bibitem[Batchelor(1952)]{batchelor-1952}
{\sc Batchelor, G.K.} 1952 Diffusion in a field of homogeneous turbulence ii.
  the relative motion of particles. {\em Proc. Cambridge Philos. Soc.\/} {\bf
  48}, 345--362.

\bibitem[Borgas \& Sawford(1994)]{borgas_etal-jfm-1994}
{\sc Borgas, M.S. \& Sawford, B.~L.} 1994 A family of stochastic models for
  two-particle dispersion in isotropic homogeneous stationary turbulence. {\em
  J. Fluid Mech.\/} {\bf 279}, 69--99.

\bibitem[Franzese \& Cassiani(2007)]{franzese_etal-jfm-2007}
{\sc Franzese, P. \& Cassiani, M.} 2007 A statistical theory of turbulent
  relative dispersion. {\em J. Fluid Mech.\/} {\bf 571}, 391--417.

\bibitem[Kolmogorov(1941)]{kolmogorov-1941}
{\sc Kolmogorov, A.N.} 1941 The local structure of turbulence in incompressible
  viscous fluid for very large {Reynolds} numbers. {\em Dokl. Akad. Nauk
  SSSR\/} {\bf 30}, 301.

\bibitem[Kurbanmuradov(1997)]{kurbanmuradov-mcma-1997}
{\sc Kurbanmuradov, O.A.} 1997 Stochastic {Lagrangian} models for two-particle
  relative dispersion in high-{Reynolds} number turbulence. {\em Monte Carlo
  Methods and Appl.\/} {\bf 3}~(1), 37--52.

\bibitem[Maurizi {\em et~al.\/}(2004)Maurizi, Pagnini \&
  Tampieri]{maurizi_etal-pre-2004}
{\sc Maurizi, A., Pagnini, G. \& Tampieri, F.} 2004 Influence of {Eulerian} and
  {Lagrangian} scales on the relative dispersion properties in {Lagrangian
  Stochastic Models} of turbulence. {\em Phys. Rev. E\/} {\bf 69}~(3),
  037301--1/4.

\bibitem[Maurizi {\em et~al.\/}(2006)Maurizi, Pagnini \&
  Tampieri]{maurizi_etal-blm-2006}
{\sc Maurizi, A., Pagnini, G. \& Tampieri, F.} 2006 Turbulence scale dependence
  of the {Richardson} constant in {Lagrangian Stochastic Models}. {\em
  Boundary-Layer Meteorol.\/} {\bf 118}, 55--68.

\bibitem[Monin \& Yaglom(1975)]{monin_yaglom-1975}
{\sc Monin, A.S. \& Yaglom, A.M.} 1975 {\em Statistical fluid mechanics\/}, ,
  vol.~II. Cambridge: MIT Press.

\bibitem[Sawford(1999)]{sawford-blm-1999}
{\sc Sawford, B.~L.} 1999 Rotation of trajectories in {Lagrangian} stochastic
  models of turbulent dispersion. {\em Boundary-Layer Meteorol.\/} {\bf 93},
  411--424.

\bibitem[Tennekes(1982)]{tennekes-1982}
{\sc Tennekes, H.} 1982 Similarity relations, scaling laws and spectral
  dynamics. In {\em Atmospheric turbulence and air pollution modeling\/} (ed.
  F.T.M. Nieuwstadt \& H.~van Dop), pp. 37--68. Reidel.

\bibitem[Thomson(1987)]{thomson-jfm-1987}
{\sc Thomson, D.J.} 1987 Criteria for the selection of stochastic models of
  particle trajectories in turbulent flows. {\em J. Fluid Mech.\/} {\bf 180},
  529--556.

\bibitem[Thomson(1990)]{thomson-jfm-1990}
{\sc Thomson, D.J.} 1990 A stochastic model for the motion of particle pairs in
  isotropic high--{Reynolds}-number turbulence, and its application to the
  problem of concentration variance. {\em J. Fluid Mech.\/} {\bf 210},
  113--153.

\end{thebibliography}

\end{document}